\documentclass[twocolumn]{aastex62}

\def\ltsima{$\; \buildrel < \over \sim\;$}
\def\ltsim{\lower.5ex\hbox{\ltsima}}
\def\gtsima{$\; \buildrel > \over\sim \;$}
\def\gtsim{\lower.5ex\hbox{\gtsima}}
\def\ms{$M_{\odot}$ }
\def\msp{$M_{\odot}$}

\received{xxx}
\revised{xxx}
\accepted{xxx}

\shorttitle{r-process feature by radial migration}
\shortauthors{Tsujimoto \& Baba}

\begin{document}

\title{Galactic r-process abundance feature shaped by radial migration}

\correspondingauthor{Takuji Tsujimoto}
\email{taku.tsujimoto@nao.ac.jp}

\author[0000-0002-9397-3658]{Takuji Tsujimoto}

\author[0000-0002-2154-8740]{Junichi Baba}

\affiliation{National Astronomical Observatory of Japan, Mitaka, Tokyo 181-8588, Japan}



\begin{abstract}
Growing interests in the chemical feature of r-process elements among nearby disk stars represented by the [Eu/Fe] vs.~[Fe/H] diagram have sprouted since it can assess the origin of r-process elements through the comparison with theoretical models, including a test as to if neutron star mergers can be the major site of r-process nucleosynthesis. On the other hand, recent studies reveal that local chemistry is strongly coupled with the dynamics of Galactic disk, which predicts that stars radially move on the disk where the observed elemental feature is different at various Galactocentric distances. Here, we show that radial migration of stars across the Galactic disk plays a crucial role in shaping the r-process abundance feature in the solar vicinity. In this proposed scenario, we highlight the importance of migration from the outer disk where [r-process/Fe] of some old stars is predicted to be enhanced to the level beyond the expectation from the observed Galactic Fe and Eu radial gradient, which results in a large span of [r-process/Fe] among nearby disk stars. The variation in the [r-process/Fe] ratio seen across the Galactic disk as well as in dwarf galaxies may be an outcome of different stellar initial mass functions which change the occurrence frequency between supernovae leaving behind neutron stars and ones ending with black holes. Here we propose that enhancement in [Eu/Fe] is attributed to the initial mass function lacking high-mass stars such as \gtsim 25\ms in the scheme for which neutron star mergers are a major source of r-process elements.
\end{abstract}

\keywords{Galaxy: abundances --- Galaxy: bulge --- Galaxy: disk --- Galaxy: kinematics and dynamics --- galaxies: dwarf --- stars: abundances}


\section{Introduction}

The elemental abundance feature in the solar vicinity represents fundamental information for understanding the origin and evolution of heavy elements \citep{Wheeler_89}. While the chemical composition of halo stars may offer a probe to assess individual nucleosynthesis events, the sequence of those of disk stars as a function of metallicity, or time shows an enrichment path built by gradual accumulation of events \citep{Tinsley_80}. This so-called Galactic chemical evolution should be grasped in the conventional framework together with a recently updated view. As an established scheme, the Galactic disk is composed of two chemically distinguishable populations, the so-called thick and thin disks \citep[e.g.,][]{Fuhrmann_98, Bensby_03, Reddy_03, Ramirez_07}, though it is still debatable whether the two disks have a separate origin or not \citep[e.g.,][]{Reddy_06, Schonrich_09a, Loebman_11, Bovy_12, Grand_18, Helmi_18}. In terms of the [$\alpha$/Fe] ratio, these two disks can be defined as high-$\alpha$ and low-$\alpha$ disk, respectively \citep[e.g.][]{Nidever_14, Hayden_17}. The two [$\alpha$/Fe]-[Fe/H] sequences are well reproduced by the models of chemical evolution with different star formation histories for the two, i.e., a short, bursty history and a long continuous one, respectively \citep[e.g.,][]{Tsujimoto_12}. Thus, disk stars are mixtures of objects that record two different paths of chemical evolution.

Recently, our knowledge of the Galactic disk has been updated by theoretical works that disk stars radially move across it due to resonant scattering with the spiral arms (and bar) \citep[e.g.,][]{Sellwood_02, Roskar_08, Schonrich_09b, Minchev_10, Baba_13, Grand_15}. This physical process implies that a treatment of chemical evolution in the solar vicinity by a simple one-zone model is insufficient and that we need to consider the contributions of stars migrating from different Galactocentric distances. In fact, it offers a nice explanation for the presence of super metal-rich stars in the solar vicinity \citep{Roskar_08}, as well as a positively skewed metallicity distribution in the outer disk \citep{Hayden_15} through radial migration from the inner disk. Moreover, interactions between the Galactic disk and satellite galaxies in the past will also induce radial mixing of stars \citep{Quillen_09, Bird_12}. Such potential disturbers of the disk are of growing importance, owing to recent evidence from stellar motions revealed by the Gaia DR2 data that the Sagittarius (Sgr) dwarf spheroidal (dSph) galaxy has perturbed the Galactic disk \citep{Gomez_13, D'Onghia_16, Schonrich_18, Kawata_18a, Antoja_18, Bland_19,Laporte_19}. 

The feature of r-process abundance among disk stars, which is represented by the [Eu/Fe] vs.~[Fe/H] diagram, has recently attracted attention. The discovery of gravitational waves from the neutron star merger (NSM) GW170817 and the subsequent discovery of multi-wavelength electromagnetic counterparts, i.e., kilonova, has identified NSMs as a major source of r-process elements \citep[e.g.,][]{Smartt_17, Pian_17, Cowperthwaite_17, Thielemann_17}. However, it has been reported that the evolutionary path of [Eu/Fe] predicted by NSMs is incompatible with the observed trend of disk stars due to a delayed r-process enrichment \citep{Hotokezaka_18}. Thus, some authors claim the necessity of r-process-element contributions from some specific and rare core-collapse supernovae (CCSNe) that are capable of enriching gas much faster than NSMs, such as magneto-rotational SNe \citep{Cote_19} or collapsars \citep{Siegel_19}. 
Similarly, stellar abundances of halo stars demand a fast enrichment of r-process \citep[e.g.,][]{Mathews_90, Argast_04}. Recent study claims that r-process enrichment by NSMs in the Galactic halo is acceptable only if a very short coalescence time ($<$1 Myr) together with neutron star creation up to a 50 \ms star is applied \citep{Matteucci_14}. It should be noted that some of the tension might be alleviated if we introduce the delay-time distribution (DTD) of lower-metallicity NSMs which is inclined to be concentrated at a shorter timescale \citep{Mennekens_16}. Current circumstances demand that the Galactic chemical evolution of r-process elements must be closely scrutinized.

\begin{figure}[t]
	\includegraphics[width=\columnwidth]{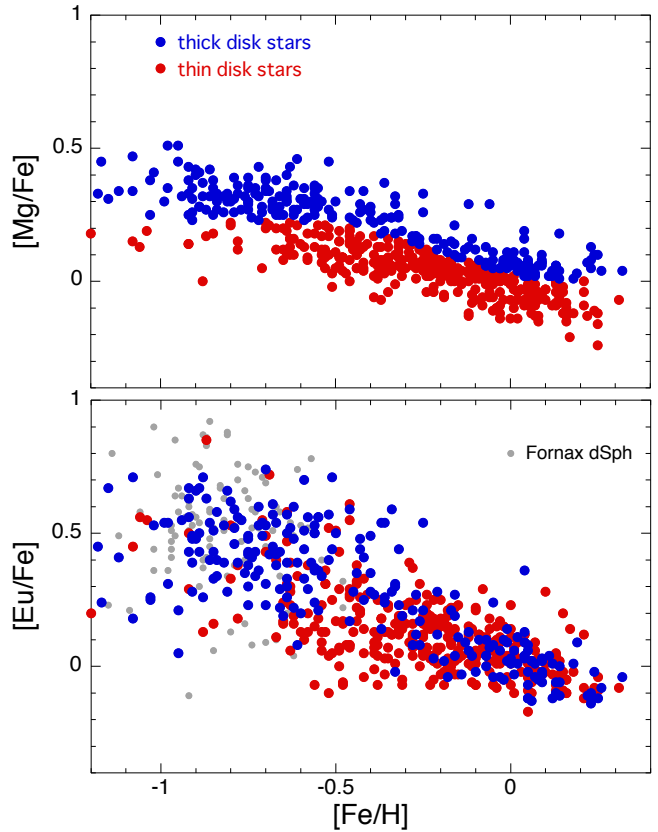}
\caption{({\it upper panel}) Observed correlation of [Mg/Fe] with [Fe/H] for disk stars in the solar vicinity. The observed data are from the SAGA database \citep{Suda_08}.The separation into two components is made in terms of individual [Mg/Fe] ratios, using the boundary between the two defined by \citet{Hayden_17}. ({\it lower panel}) The [Eu/Fe] features for the thick and thin disks, which correspond to high-[Mg/Fe] and low-[Mg/Fe] stars in the upper panel. Observed [Eu/Fe] ratios of stars in the Fornax dSph \citep{Letarte_10, Lemasle_14} are also attached with gray circles.}
\end{figure}

As discussed, the first step toward implementing this task would be to see the [Eu/Fe] trends for thick and thin disks separately. The upper panel of Figure 1 shows the [Mg/Fe] features for the two disks, using the  boundary between the high- and low-$\alpha$ regions adopted by \citet{Hayden_17}. Then, we obtain the corresponding [Eu/Fe] feature for each component (the lower panel). We see that (i) the thick disk has higher [Eu/Fe] at the same [Fe/H], and that (ii) there are metal-poor stars with extremely high [Eu/Fe] values such as [Eu/Fe]\gtsim$+0.7$. Since the mean values of [Eu/Fe] for metal-poor stars of both disks are higher than those of [Mg/Fe], a contrast in [Eu/Fe] between metal-poor and metal-rich stars is larger than that of [Mg/Fe]. A similar result is also obtained from a different database \citep{Guiglion_18}. In addition, the two disks separated in terms of age also share a similar range of [Eu/Fe] distribution such as $-0.1$\ltsim [Eu/Fe] \ltsim$+0.6$ \citep{Battistini_16}, which may alleviate concern about the fact that our approach to separate into the two disks suffers an inadequate selection caused by an error in [Mg/Fe]. A large scatter in [Eu/Fe] may be, in part, caused by an observational uncertainty of Eu abundance which is deduced from the only one line in most cases. However, it is unlikely that this could be the main reason for the spread in [Eu/Fe] since a scatter in [Eu/Fe] among metal-poor disk stars is larger than that of a plateau made by halos stars around [Fe/H] =  $-1.5$ by about 0.3 dex (e.g., the SAGA database), as opposed to our expectation that halo stars should exhibit a larger scatter as an outcome of stochastic enrichment at an early phase of halo formation. In the end, the revealed feature suggest a large difference in [r-process/Fe] between metal-poor and metal-rich disk stars, which challenges the enrichment of Eu by NSMs. That's because r-process enrichment with the plausible DTD spanning over Gyrs for NSMs predicts only a small gradient in [Eu/Fe] with respect to [Fe/H] between the two endpoints of chemical evolution, compared to the case with fast r-process enrichment by CCSNe \citep{Hotokezaka_18, Cote_19, Siegel_19}.

The radial-migration scheme directs us to regard the solar vicinity as one piece of the Galactic disk. Then,  we notice that there is a positive [Eu/Fe] gradient across the Galactic disk, which is observationally identified by open clusters \citep{Yong_12, Overbeek_16} and Cepheids \citep{daSilva_16, Luck_18}. Therefore, the mixture of stars migrating from various Galactocentric distances has the potential to explain the large scatter in [Eu/Fe], as observed in the solar vicinity. This view may be supported by the observed [Eu/Fe] feature in the Galactic bulge where the effect of radial migration is not expected: there are no stars with extremely high [Eu/Fe] values \citep[][see Fig.~2]{Johnson_12, Barbuy_14, Siqueira_16}. Besides the outer disk, we see high [Eu/Fe] ratios in the Milky Way satellites such as the Fornax dSph galaxy (gray circles in Fig.~1), which may imply the possible connection of an enhanced [Eu/Fe] ratio with a low-metallicity environment owing to a low star formation rate. Here we note that the [Eu/Fe] distribution of the Fornax dSph is very similar to that of the Galactic thick disk. However, it remains unclear whether the presence of stars with very high [Eu/Fe] values is a common feature among nearby dSphs: the Carina dSph shows a large span of [Eu/Fe] including [Eu/Fe]$\approx+1$ \citep{Norris_17} while there are no extreme [Eu/Fe] values in the Ursa Minor dSph \citep{Mashonkina_17}.

This connection invokes an idea that the initial mass function (IMF) variations may be one way to explain enhancement in [Eu/Fe]. Note that stochastic chemical evolution models can produce high [Eu/Fe] values only for very metal-poor stars with [Fe/H]\ltsim$-2$ \citep{Cescutti_15, Hirai_15, Hirai_17}. The mechanism for enhancing this ratio is that r-process production is associated with only low-mass CCSNe which leave neutron stars while all CCSNe produce Fe, and that the IMF with a low high-mass end results in lifting the [r-process/Fe] ratio. Indeed, the suppression of the formation of massive stars in low surface brightness galaxies is implied from the observed flux ratio of H$_\alpha$ to the far-ultraviolet \citep{Meurer_09, Pflamm_09, Lee_09}. Such an implied  IMF lacking high-mass stars could enhance [Eu/Fe] sufficiently to cover all [Eu/Fe] range by nearby disk stars and must be realized in the outer disk as in dwarf galaxies. As a supporting discussion, enhanced [Eu/Mg] in the Fornax dSph is claimed to be a result of the IMF that lacks the most massive stars \citep{Lemasle_14}.

\section{Chemical evolutions of closed and open systems in the Galaxy}

\begin{figure}
	\includegraphics[width=\columnwidth]{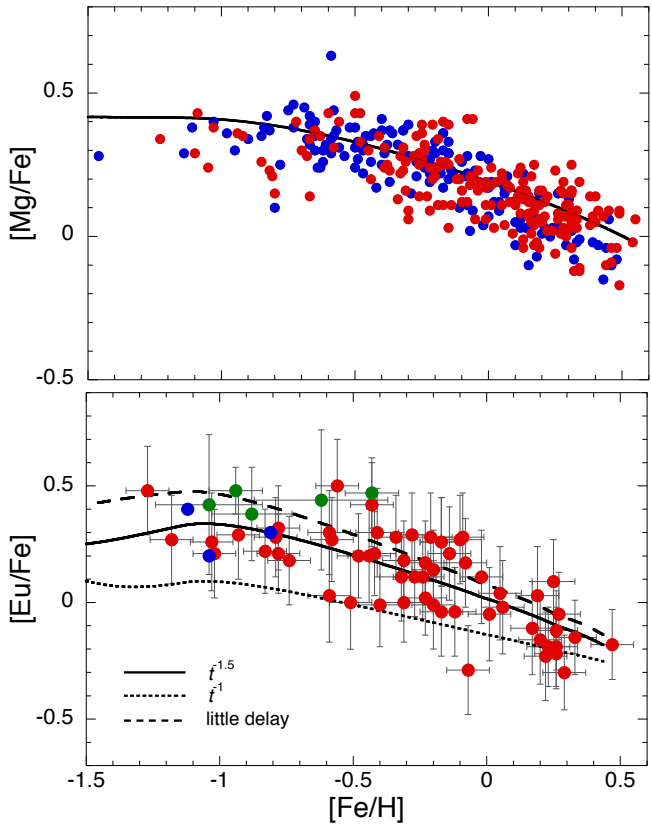}
    \caption{Chemical evolution of the Galactic bulge. ({\it upper panel}) Calculated [Mg/Fe] evolution by the one-zone chemical evolution model, compared with the observed data (blue circles: Gonzalez et al.~2011; red circles: Johnson et al.~2014). ({\it lower panel}) Same as the top panel, but for [Eu/Fe]. We calculate three cases with different delay-time distributions of r-process (Eu) enrichment: $\propto t^{-1.5}$ (solid line), $\propto t^{-1}$ (dotted line), and little delay-time of 10-30 Myr (dashed line). For the former two cases, the minimum delay time is set to be 10 Myr. The observed data are taken from Johnson et al.~2012 (red circles); Barbuy et al.~2014 (blue circles); Siqueira et al.~2016 (green circles).}
\end{figure}

First, we examine the chemical evolution of two components in the Galaxy. The first is the Galactic bulge. Since we do not expect any radial-migration effects there, chemical evolution of the bulge should be basically reproduced by a one-zone model. Thus, we can regard the bulge as a closed system in terms of the absence of stellar migration, and give a stringent constraint on the DTD of r-process producer which is a matter of controversy by applying one-zone chemical evolution models for the bulge. The other is for the Galactic disk, which is open to stellar migration. As a result, one-zone models for the disk equipped with the proper DTD may explain only part of the observed features of the elements whose abundances differ from [Fe/H] in terms of their dependence on Galactocentric distances. Here, we focus upon the evolutions of [Mg/Fe] and [Eu/Fe] with respect to [Fe/H].

\subsection{Galactic Bulge}

Chemical evolution of the bulge can be reproduced in the scheme of a relatively rapid star formation. We adopt the same model as in \citet{Tsujimoto_12}: a short timescale of star formation ($\tau_{\rm SF}$ = 0.5 Gyr) for the duration of $\Delta_{\rm SF}$ = 2 Gyr with a rapid collapse ($\tau_{\rm in}$ = 0.3 Gyr). For the DTD of Type Ia SNe (SNe Ia), we adopt DTD $\propto t_{\rm delay}^{-1}$ with a range of $0.1\leq t_{\rm delay}\leq10$ Gyr \citep{Hachisu_08, Totani_08}. The DTD is normalized so that 8\% of the primary stars in binaries with initial masses in the range of $3-8$\ms explode as SNe Ia. Then, as shown in the top panel of Figure 2, we see that the calculated [Mg/Fe] evolution is compatible with the observed data.

To calculate the Eu abundance, we assume that r-process elements are produced in NSMs. We adopt a Eu mass of $8.3\times 10^{-5}$\ms ejected from a single NSM by assuming that the ejecta of an NSM have a mass of 0.01 \ms \citep[e.g.,][]{Barnes_13} and comprise elements with $A\geq 127$, as is implied from meteoritic abundances \citep{Tsujimoto_17}. In addition, NSM events are assumed to occur at a rate of one per 1,000 CCSNe \citep{Tsujimoto_14}. For the DTD of NSMs, we consider three different cases. In the first case, DTD is assumed to be proportional to $t_{\rm delay}^{-1}$ with a range of $0.01\leq t_{ \rm delay}\leq10$ Gyr. This DTD form of $t_{\rm delay}^{-1}$ is implied from that for short gamma-ray bursts \citep{Fong_17}. The second DTD has a steeper slope of $-1.5$ to be weighted at an early emergence of NSMs, compared to the first case, which is motivated by previous reports \citep{Hotokezaka_18, Cote_19}. Moreover, as an extreme case, we prepare a model with little time delay, i.e., DTD which is restricted within 10-30 Myr as adopted by \citet{Tsujimoto_14}. The third case may be identical to that based on the hypothesis that the major source of r-process elements is CCSNe such as collapsars \citep{Siegel_19}.
 
The results are shown in the bottom panel of Figure 2. As pointed out by previous studies on the Galactic disk, the model with DTD $\propto t_{\rm delay}^{-1}$ is incompatible with the observed features of the Galactic bulge in the same way. Alternatively, the models which are capable of faster enrichment in Eu as exemplified by the other two DTDs can reproduce the observations. Therefore, without uncertainties in the contaminant from outside, our result leads us to the conclusion that the Eu abundance feature in the bulge supports the faster ejection of r-process elements than the delay expected from the DTD (i.e., $\propto t_{\rm delay}^{-1}$) of short gamma-ray bursts. This may imply two possibilities of r-process production site; a sole site of NSMs with a steeper DTD or a mixture of sites including rare CCSNe.

\subsection{Galactic Disks}

By incorporating an ingredient for r-process enrichment $-$the DTD form of $\propto t_{\rm delay}^{-1.5}$ in a scheme that a NSM is the major site of r-process production $-$ into the one-zone models, we examine the chemical evolution of the Galactic disk in the solar vicinity. For the thick disk, we assume a rapid formation, similar to the case of the Galactic bulge, as modeled by ($\tau_{\rm SF}$, $\Delta_{\rm SF}$, $\tau_{\rm in}$)=(0.25, 1, 0.3) in units of Gyr. On the other hand, the thin disk is assumed to be gradually formed with inputs ($\tau_{\rm SF}$, $\Delta_{\rm SF}$, $\tau_{\rm in}$) = (1.5, 10, 5). We connect the formation of the two disks so that the thin disk was formed after the formation of the thick disk and thus the thin-disk stars started forming from the thick disk's remaining gas mixed with the gas accreted onto the disk. Accordingly, the chemical abundances of the first stars in the thin disk are similar to those of the most metal-rich stars in the thick disk. Details of the model description can be found in \citet{Tsujimoto_12}. 

\begin{figure}
	\includegraphics[width=\columnwidth]{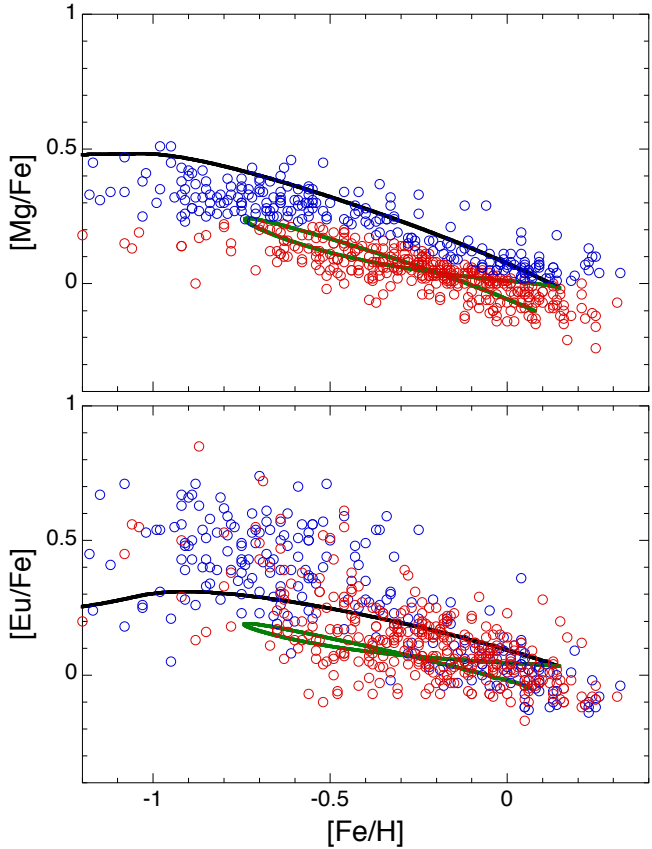}
    \caption{Chemical evolution of the Galactic disk predicted by one-zone models. The disk is modeled separately for the thick and thin disks. ({\it upper panel}) Comparison of the calculated [Mg/Fe] with the observations. The observed data are the same as in Figure 1. The model result for the thick disk is shown by a black line, while the green line pertains to the thin disk. ({\it lower panel}) Same as the top panel, but for [Eu/Fe].}
\end{figure}

As shown in the top panel of Figure 3, the thick-disk formation leaves the metal-rich gas as an end product of its chemical evolution. Thus, the evolution of [Mg/Fe] in the thin disk starts from [Fe/H] $\sim +0.15$. Then, [Fe/H] and [Mg/Fe] decrease and increase, respectively, owing to dilution by metal-poor infalling gas. This reverse evolution comes to an end when the chemical enrichment by star formation exceeds the effect of gas dilution, and subsequently a usual evolutionary path appears. We see that the predicted [Mg/Fe] ratios for both disks well explain the distribution of the observed data. On the other hand, this is not the case for [Eu/Fe] (the bottom panel in Fig.~3). The [Eu/Fe] ratios predicted by the one-zone model occupy only the lower part of the observed distribution.  In addition, as already mentioned, there exists a radial gradient of [Eu/Fe] in the sense that [Eu/Fe] becomes higher at larger Galactocentric distances \citep{Yong_12, Overbeek_16, daSilva_16}. These results point to the possible origin of stars with high [Eu/Fe] ratios beyond the prediction by one-zone models as migration from the outer disk. 

\begin{figure}
	\includegraphics[width=\columnwidth]{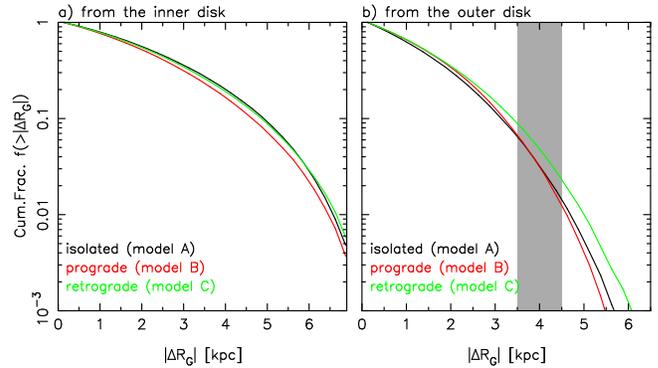}
    \caption{Cumulative fraction of stars from the inner disk (left) and the outer disk (right) to the solar vicinity as a function of the distance from the sun predicted by the three simulation models. The final result after the 7.5 Gyr-evolution are shown. The models represent cases without any effects by satellite galaxies (isolated) and ones including interaction with a satellite galaxy as massive as the Sgr dSph galaxy with prograde and retrograde planar orbits. The shaded regions in the right panel indicate the outer disk where the [Eu/Fe] ratios of old stars are observationally implied to be sufficiently high to explain the upper part of the [Eu/Fe] range exhibited by nearby disk stars.}
\end{figure}

\section{Simulations with radial migration}

\begin{figure*}
\begin{center}
	\includegraphics[width=2.0\columnwidth]{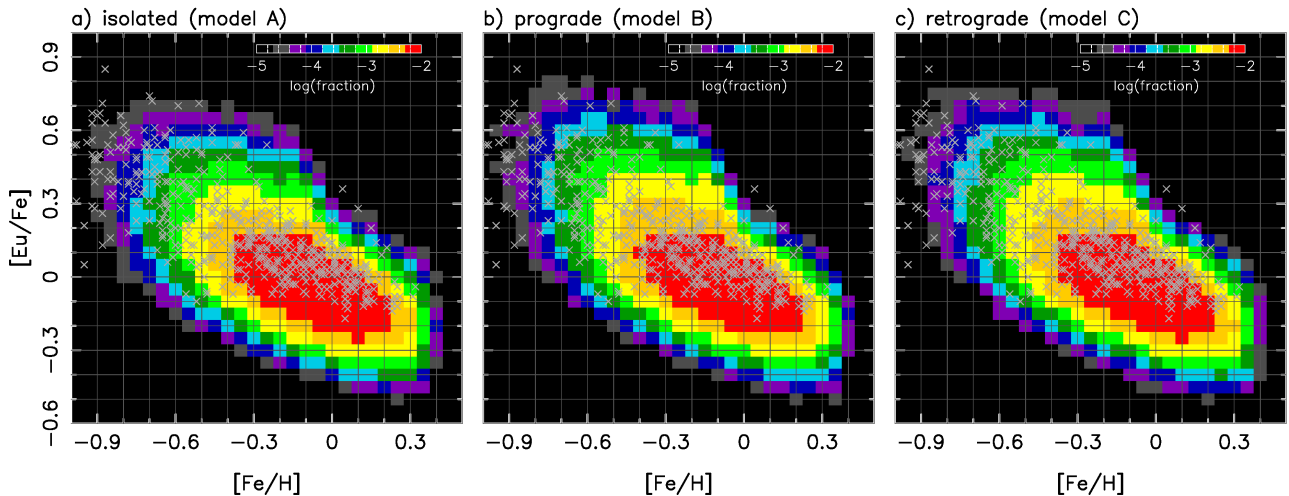}
\end{center}
    \caption{The predicted [Eu/Fe] vs.~[Fe/H] diagrams as a result of radial migration over the 7.5 Gyr-evolution by the three simulation models, as compared with the observed data \citep[crosses:][]{Suda_08}. Colors indicate the normalized fraction in logarithmic scale.}
\end{figure*}

In order to investigate the effect of stellar radial migration upon the chemical evolution of the disk, we perform idealized $N$-body simulations of Milky Way-like disk galaxies using the ASURA code \citep{Saitoh_08, Saitoh_13, Saitoh_17}. We generate the initial axisymmetric model of a Milky Way-like stellar disc with a classical bulge embedded in a dark matter (DM) halo. The stellar disk follows an exponential profile with a mass of $4.5 \times 10^{10}$ \msp, a scale-length $R_{\rm d}$ of 2.6 kpc, and a scale-height of 300 pc. Using Hernquist's method \citep{Hernquist_93}, the velocity structure of the stellar disk in cylindrical coordinates is approximately determined by a Maxwellian approximation. We set the reference radial velocity dispersion by assuming that Toomre's $Q$ at $R_{\rm G}$ = 2.5 $R_{\rm d}$ equals 1.1. In this study, for simplicity, we assume the classical bulge and the DM halo to be a static potential, whose density profiles follow the Hernquist profile \citep{Hernquist_90} and the Navarro-Frenk-While profile, respectively. Here, the mass and scale-length of the classical bulge are set to be $4.3 \times 10^9$ \ms and 350 pc, respectively. For the DM halo, we assume that the mass and concentration parameter are $9.4 \times 10^{11}$ \ms and 13.7, respectively. Our models contain 10 million particles. We use a gravitational softening length of 10 pc. More detailed model description can be found in \citet{Baba_15}.

Here, we consider three $N$-body simulations of Milky Way-like disk galaxies: an isolated disk model (model A) and two perturbed disk models driven by the interaction with Sgr dwarf-sized satellite galaxies, which are treated as test particles and are assumed to have prograde (model B) and retrograde (model C) planar orbits, respectively. The modeled satellite has the mass of $9\times 10^9$\ms with the pericenter around 20 kpc. All three models form a stellar bar with a semi-major length of about 3 kpc and spiral arms. In this study, we do not show the result of the actual Sgr case since the interaction with the Sgr dSph is found to have little effect upon radial migration, owing to its nearly polar orbit which induces only a little redistribution of the angular momentums of disk stars.

Next, we paint N-body models with abundances (i.e., [Fe/H] and [Eu/Fe]). This painting technique has been used to investigate the evolution of metallicity gradient or relationships between chemistry and kinematics of galaxy disks by many studies \citep[e.g.,][]{Curir_12, Matteo_13,Grand_15, Kawata_18b}, though this technique, depending on the situation, confronts a difficult task that the star formation history has to be strictly coupled to chemical evolution \citep{Schonrich_17}, particularly at the ealy star formation \citep{Minchev_14}. In this study, we take a simple approach: we set all stars initially and do not consider the time-variation in elemental abundances of [Fe/H] and [Eu/Fe] for the overall disk. We assume that individual stars within the solar circle from 8$<R_{\rm G}<9$ kpc are chemically tagged with a set of ([Fe/H], [Eu/Fe]) so as to match with the chemical features in the solar vicinity modeled in \S 2.2 as follows. The values of [Fe/H] are assigned to individual stars to reproduce the predicted [Fe/H] distribution function of thin disk stars, which ranges for $-0.8\leq$[Fe/H]$\leq +0.1$ with its peak at [Fe/H]=$-0.15$ and is compatible with the observed one \citep{Bekki_11}. Subsequently, each star painted with [Fe/H] is given a [Eu/Fe] value in accordance with a correlation of [Eu/Fe]-[Fe/H] predicted by the thin disk model. Here we adopt the evolutionary path starting from [Fe/H]=$-0.8$ with [Eu/Fe]=+0.2 and ending at [Fe/H]=+0.05 with [Eu/Fe]=$-0.05$ since the majority of stars are on this track.
Then, we give the final distributions of [Fe/H] and [Eu/Fe], considering their dispersions of 0.2 and 0.1 dex, respectively.

\begin{figure*}
\begin{center}
	\includegraphics[width=2.0\columnwidth]{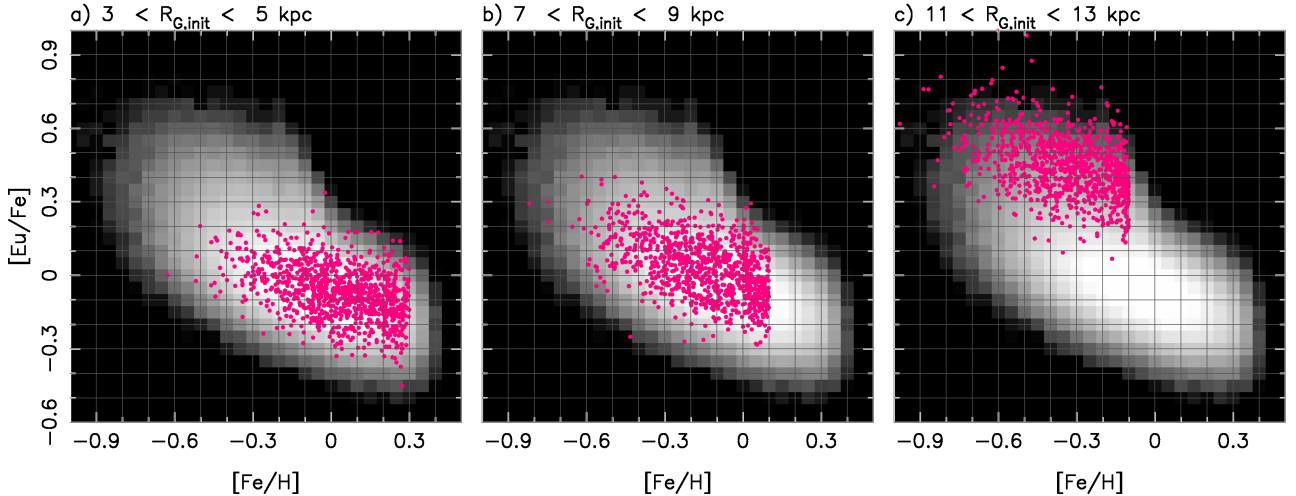}
\end{center}
    \caption{The modeled [Eu/Fe]$-$[Fe/H] distribution of stars at three different Galactocentric distances (magenta circles), which are overlaid on the final [Eu/Fe]$-$[Fe/H] plane in the solar vicinity obtained for the isolated case (model A) of Figure 5 (grey region). Each distribution contains 1000 stellar particles which are randomly selected.}
\end{figure*}

For other disk regions at different Galactocentric distances, each collection of stars is assumed to be analogous to the above with its mean [Fe/H] and [Eu/Fe] shifted by the amount $\Delta$[Fe/H] ($\Delta$[Eu/Fe]) expected from Galactic-radial [Fe/H] ([Eu/Fe]) gradient. Namely, at each Galactocentric radius, we set a group of stars in which their [Fe/H] distribution is shifted by $\Delta$[Fe/H] as a whole and their [Eu/Fe] values are determined by a shift of $\Delta$[Eu/Fe] from a value deduced from the [Eu/Fe]-[Fe/H] correlation at the solar circle. For the [Fe/H] gradient, we adopt a slope of $-0.05$ dex kpc$^{-1}$ \citep[e.g.,][]{Genovali_14}. As for the [Eu/Fe] gradient, we adopt the value of +0.125 dex kpc$^{-1}$ for the outer disk which is broadly equivalent to the maximum limit $\Delta$[Eu/Fe]$\approx 0.5$ dex of its gradient  between $R_{\rm G}$ = 8.5 and $\sim$13 kpc in the data of \citet{Yong_12}. Note that the mean [Eu/Fe] gradient for the corresponding region is deduced to be +0.07 dex kpc$^{-1}$ \citep[][and also +0.039 dex kpc$^{-1}$ for the wider region by Oberbeek et al.~2016]{Yong_12}, and that the assumed steep [Eu/Fe] gradient results in the positive [Eu/H] gradient of +0.075 dex kpc$^{-1}$ which is in contradiction with observational data from open clusters and Cepheids \citep[e.g.,][]{Luck_18}. We anticipate that such a steep [Eu/Fe] gradient and thus a positive [Eu/H] gradient should be present for old stars in the outer disk. On the other hand, a shallower slope of +0.02 dex kpc$^{-1}$ for the inner disk is assigned, taking into account an implication from the lower bound of [Eu/Fe] at the innermost part, i.e., the Galactic bulge. 

The feature of migration from both the inner disk and the outer disk predicted by our models (A-C) is summarized in Figure 4. Here we show the model result corresponding to the present day. Our isolated model result is in good agreement with those by previous studies \citep[e.g.,][]{Schonrich_09b, Loebman_11, Aumer_17}: for instance, the stars migrating from the inner disk to the solar vicinity are a few times of those from the outer disk. Note, however, that our models may predict migration from further outer disk up to $\Delta R_{\rm G}\approx$ 6 kpc, compared to $\sim$4 kpc by other studies. As a result of the long-term evolution, all models eventually predict migration of outer disk stars to the solar circle from the Galactocentric distances (shaded region) where old stars are expected to be exhibit [Eu/Fe] ratios as high as $\sim+0.5\pm 0.2$, as observed for some fraction of nearby disk stars. Among our three models, the one including the satellite with a retrograde orbit (model C) is found to induce the most efficient migration from the outer disk. It should be noted that the outer disk resulting from model B shows a more perturbed morphology than that of model C. Figure 5 shows the resultant [Eu/Fe] vs.~[Fe/H] diagrams in the solar vicinity according to the three models, as compared with the observed data. We see a good agreement between the model results and the observations: a large span in [Eu/Fe] of nearly 1 dex is eventually reproduced  owing to radial migration. As seen in the similarity of three results, we find little significant effect of radial migration driven by the potential interaction with a satellite galaxy on the final [Eu/Fe] feature. Figure 6 demonstrate how the modeled [Eu/Fe]-[Fe/H] planes at different Galactocentric distances put together into the final abundance plot in the solar vicinity for the case of model A. It reveals that stars with [Eu/Fe]\gtsim$+0.3$ owe their origin to the outer disk, and that stars with [Fe/H]\gtsim 0 with low [Eu/Fe] (\ltsim 0) are migrated from the inner disk. According to \citet{Battistini_16}, there might be a weak age-[Eu/Fe] correlation in the sense that stars with high [Eu/Fe] are old. This can be compatible with our models if age distributions of stars across the Galactic disk are similar \citep{Hayden_15}. 

\section{The origin of high [r-process/Fe] ratio}

Finally, we discuss the possible origin of the enhanced [r-process/Fe] ratio in the outer disk. We anticipate that old stars formed before a contribution to Fe from SNe Ia commenced in the outer disk exhibit a high [Eu/Fe] ratio such as $\sim+0.5\pm0.2$. We suggest to invoke some ingredient for the chemical evolution which is different from that for the solar vicinity. In this argument, we add one observed fact: the plateau of [Eu/Fe] among nearby halo stars in the metallicity range of $-2\leq$[Fe/H]$\leq-1.5$ shows [Eu/Fe] $\approx +0.4$ on average, which leads to an offset in [Eu/Fe] between its halo plateau and metal-poor nearby disk stars. Here, we could recall that r-process production is connected to low-mass CCSNe, which leave behind NSs in the current scheme for which  NSMs are the major source of r-process elements. The boundary of the mass of progenitor stars that collapse to form a NS or a black hole lies around 20-25 \ms \citep{Fryer_99}. This boundary mass is also supported by recent detailed supernova explosion models, giving $\sim$25\ms as the most plausible value \citep[][see also Muller et al.~2016]{Ebinger_19}. While stars with masses larger than 20-25 \ms are not therefore associated with r-process production, they contribute to Fe production in a sufficient way according to a larger Fe yield from a more massive star up to $\sim 50$\ms \citep{Tominaga_07}. Suppose that a certain galaxy lacks massive stars with $\geq$25 \msp. Then, Fe production will reduce to about one third of the yield provided by a full range of CCSNe, while r-process production remains unchanged. These outcomes eventually lead to an increase in [Eu/Fe] by $\sim 0.5$ dex. 

Indeed, for instance, the lack of very massive (\gtsim 25 \msp) stars is suggested for relatively massive dSph galaxies such as the Fornax dSph based on s-process elements \citep{Tsujimoto_11} and the Sgr dSph from $\alpha$-elements \citep{McWilliam_13}. Figure 1 shows that the Fornax exhibits high [Eu/Fe] ratios up to +0.9. The Sgr stars also show a high [Eu/Fe] of $\sim+0.5$ around [Fe/H] = 0 while the [$\alpha$/Fe] ratios are below the solar ratio \citep{McWilliam_13}. In addition, there is observational evidence for a nonuniform IMF suggesting that a lower (higher) star formation matches with a smaller (larger) number of massive stars \citep[][see also Pflamm-Altenburg et al.~2009; Lee et al.~2009]{Meurer_09}. Their results suggest that low star-formation systems possess the IMF with a high-mass end around 30 \msp, which would yield few CCSNe leaving black holes. Thus, we can anticipate star formation lacking very massive stars which leads to the sufficiently enhanced [Eu/Fe] ratio in the outer disk. On the other hand, a slightly lower [Eu/Fe] in the inner disk may be attributed to an IMF favoring very massive stars collapsing to black holes, owing to a higher star formation rate. 

Then, we can connect the IMF variation to the assumed [Eu/Fe] gradient, for instance, as follows. The combination of the Fe yield \citep{Shigeyama_98, Tominaga_07} with a change in the high-mass end from 50 \ms to 25 \ms gives $\Delta$[Eu/Fe]$\approx$ +0.4-0.5. If we set the IMF with such a cut-off in the outer disk around $R_{\rm G}$ = 12.5 kpc, the [Eu/Fe] slope could be 0.1 - 0.125 dex kpc$^{-1}$. On the other hand, a higher star formation (e.g., $\tau_{\rm SF}$ = 1 Gyr) with a flatter IMF with a slope $x$ of $-1.2$ can increase the current [Fe/H] by $\sim$0.25 dex, compared with the case for the solar vicinity holding the Salpeter IMF ($x$=$-1.35$). Such a flatter IMF predict a small decrease in [Eu/Fe] of $-0.05$ - $-0.1$, taking into account an increase in the number of NSs. If we set this IMF at $R_{\rm G}$ = 3.5 kpc where a difference in [Fe/H] by 0.25 dex from the solar circle is expected, the [Eu/Fe] slope could be 0.01 - 0.02 dex kpc$^{-1}$.

\section{Conclusions}

We show that the presence of nearby disk stars with enhanced [Eu/Fe] ratios such as \gtsim +0.4 are outside the prediction from one-zone chemical evolution models and conclude that they migrated from the outer disk owing to the interactions with spiral arms over the long-term evolution of Galactic disk. In our models, we assume that the [Eu/Fe] gradient in the outer disk has a steep slope as large as +0.125 dex kpc$^{-1}$, which is steeper than the observed gradients of intermediate-age (open clusters) and young (Cepheids) objects. This assumed [Eu/Fe] gradient yields a positive radial [Eu/H] gradient which is not supported by the current observational data; however, it is anticipated to be realized for old stars. Even if the outer disk did not have such a steep gradient initially, a mildly enhanced [Eu/Fe] as observed in the outer disk must contribute, in part, to the r-process abundance feature in the solar vicinity through radial migration.

Our scenario predict that very old stars in the outer disk should show a large enhancement in [Eu/Fe] as high as those seen in some dwarf galaxies such as the Fornax dSph, and that there is a difference in the upper bound of [Eu/Fe] between the inner disk and the outer disk. On the other hand, a long-term chemical evolution would result in a mildly enhanced [Eu/Fe] at the present outer disk as suggested by abundances of young objects such as Cepheids. Then, the origin of enhancement in [Eu/Fe] is discussed in the scheme for which NSMs are the major site of r-process production. We propose that an enhanced [Eu/Fe] is an end result of the IMF lacking high-mass stars, which reduces Fe production in total while keeping almost the same for the amount of r-process elements \citep[see, however, another possible scenario presented by][]{Schonrich_19}.

The potential IMF variation with Galactocentric radius can be interpreted to be a natural result of mass differences in star-forming molecular clouds. It is observationally found that the inner disk host the most massive molecular clouds and their average mass decreases with Galactocentric distance \citep{Miville_17}. On the other hand, theoretical considerations reveal that the mass of star cluster formed from collapsing molecular clouds increases in accordance with the mass of molecular clouds \citep{Howard_18}, and that less (more) massive star clusters possess a smaller (larger) value of high-mass end of the IMF \citep{Kroupa_03, Weidner_05, Pflamm_08}. Putting it all together, we predict that the IMF varies with Galactocentric radius, which should be imprinted on the variation in [Eu/Fe] values across the Galactic disk. 

Our proposed IMF variations in the Galactic disk do not have a large impact on the [Mg/Fe] feature. This is attributed to a similar dependence of Mg and Fe productions on the mass of progenitor stars. Here, the Mg yield is deduced from nucleosynthesis calculations while the Fe yield is estimated from the analyses of a light curve of supernovae or elemental abundances of very metal-poor stars \citep[see][]{Shigeyama_98}. Based on this argument, the variation in the IMF would hardly change the [Mg/Fe] ratio. In the end, the [Mg/Fe] feature across the Galactic disk is set up by either the inside-out scenario \citep{Chiappini_01, Cescutti_07} or radial gas flows \citep[e.g.,][]{Gotz_92, Tsujimoto_95, Portinari_00, Pezzulli_16}, both of which predict a very small [Mg/Fe] gradient owing to similar negative [Mg/H] and [Fe/H] gradients, regardless of whether the IMF variations exist or not. Accordingly, the [Mg/Fe] feature could be  characterized by a similar range of [Mg/Fe] distribution at various Galactocentric distances, which is observationally supported by the APOGEE data \citep{Hayden_15}. A small positive radial gradient of [$\alpha$/Fe] implied by Cepheids \citep[e.g., $\sim$+0.015 dex kpc$^{-1}$ for Mg:][]{Genovali_15} could be interpreted by theoretical prediction that delayed chemical evolution in the outer disk caused a relatively small contribution of Fe from SNe Ia to stellar abundances, compared with those by CCSNe \citep{Cescutti_07}.

\acknowledgements

We would like to thank our referee, Ralph Sch\"{o}nrich, whose detailed and accurate comments much improved the presentation of our paper. TT also thanks Takuma Suda for his advice on practical use of the SAGA database. This work was supported by JSPS KAKENHI Grant Numbers 18H01258, 18H04593, 16H04081, 18H01248, 17H02870, and 18K03711.




\end{document}